\documentclass[]{article}
\usepackage[utf8]{inputenc}
\usepackage[margin=1in]{geometry}
\usepackage{amsmath}
\usepackage{amssymb}
\usepackage{graphicx}
\usepackage{authblk}
\usepackage{ulem}
\usepackage{hyperref}
\usepackage{xcolor}

\title{\textbf{Biophysics underlying the swarm to biofilm transition}}
\author[1]{Vasco M. Worlitzer}
\author[2]{Ajesh Jose}
\author[3]{Ilana Grinberg}
\author[1]{Markus Bär}
\author[1]{Sebastian Heidenreich}
\author[3]{Avigdor Eldar}
\author[4]{Gil Ariel}
\author[2,5]{Avraham Be'er}
\affil[1]{Department of Mathematical Modelling and Data Analysis, Physikalisch-Technische Bundesanstalt Braunschweig und Berlin, Abbestrasse 2-12, D-10587 Berlin, Germany}
\affil[2]{Zuckerberg Institute for Water Research, The Jacob Blaustein Institutes for Desert Research, Ben-Gurion University of the Negev, Sede Boqer Campus, 84990 Midreshet Ben-Gurion, Israel}
\affil[3]{The Shmunis School of Biomedicine and Cancer Research, Faculty of Life Sciences, Tel Aviv University, Israel}
\affil[4]{Department of Mathematics, Bar-Ilan University, 52900 Ramat Gan, Israel}
\affil[5]{Department of Physics, Ben-Gurion University of the Negev, 84105 Beer Sheva, Israel}

\begin{document}

\maketitle

\begin{abstract}
    Bacteria organize in a variety of collective states, from swarming, which has been attributed to rapid surface exploration, to biofilms, which are highly dense immobile communities attributed to stress resistance. It has been suggested that biofilm and swarming are oppositely controlled, making this transition particularly interesting for understanding the ability of bacterial colonies to adapt to challenging environments. Here, the swarm to biofilm transition is studied experimentally by analyzing the bacterial dynamics both on the individual and collective scales. We show that both biological and physical processes facilitate the transition - a few individual cells that initiate the biofilm program cause nucleation of large, scale-free stationary aggregates of trapped swarm cells. Around aggregates, cells continue swarming almost unobstructed, while inside, trapped cells slowly transform to biofilm. While our experimental findings rule out previously suggested purely physical effects as a trigger for biofilm formation, they show how physical processes, such as clustering and jamming, accelerate biofilm formation.
\end{abstract}

\section{Introduction}
Bacteria organize in a variety of collective states, which are referred to as bacterial lifestyles. Transitions between lifestyles, in which entire bacterial populations change their properties – both at the individual and colony-wide collective levels -- are quintessential examples of adaptation to environmental signals and stresses \cite{Kolter2006, Kearns2005, Lopez2010, Desai2019, Srinivasan2019, Grobas2020}. Biofilms are one of the best studied lifestyles, in which multicellular aggregates of bacteria are bound, both to each other and to surfaces, by an extracellular polymeric substances (EPS) that they secrete \cite{Hall-Stoodley2004a, Vlamakis2013}. Formation of biofilm can occur under different conditions. Correspondingly, transitions to a biofilm lifestyle may take place starting from a variety of other lifestyles such as planktonic (free-floating) bacteria or dense colonies, and from a variety of environments e.g., liquid bulk, surface-attached cell, wet habitats or dry ones. Here, we consider the less-studied transition from an active swarm to a biofilm. Bacterial swarming is a collective mode of motion in which cells move rapidly over surfaces, forming dynamic patterns of whirls and jets \cite{Kearns2003, Kearns2010, Beer2019}. It has been suggested that biofilm and motility, specifically swarming, are oppositely controlled \cite{Guttenplan2013}, making this transition particularly interesting.

Transitions into a biofilm involve a range of cellular changes such as changes in gene expression, motility regulation and secretions of EPS \cite{Hall-Stoodley2004a, Guttenplan2013}. The stages of these biological processes, in which bacteria switch states, can be referred to as a biofilm program \cite{Desai2019, Hall-Stoodley2004a, Vlamakis2013}. At the same time, the changing mechanical properties of cells and their environment impose new physical constraints on cells, leading to new dynamical patterns and possibly phase transitions (in the statistical physics sense) \cite{Jeckel2019, Beer2020}. The intricate interactions between the biological process involved in the biofilm program and the physics of the bacterial collectives are largely unknown. In particular, it is not clear which effect precedes the other \cite{Guttenplan2013}. It is known that production of EPS results in cell adhesion and eventual immobility of all cells. From this perspective, a biological process leads to a change in mechanical cell-properties and new physical conditions. On the other hand, it has been suggested that high cell densities lead to jamming and possibly phase separation \cite{Grobas2021}. Physical theories, for example the well-studied motility-induced phase separation (MIPS), predict that cells may self-segregate into two phases: a low-density phase of freely moving bacteria (an "active gas" of self-propelled particles), and high-density aggregates of stuck and immobile (jammed) bacteria \cite{Fily2012, Bialke2013, Cates2015}. In turn, jammed cells that are temporarily stuck, experience torque that resists flagellar rotation, which may initiate the biofilm program \cite{Cairns2013}. This point of view suggests a reversed story, in which a physical phenomenon, namely jamming and phase separation, triggers the biological one -- biofilm program. 

\textit{Bacillus subtilis} is a major model organism for studying both biofilm formation and swarming \cite{Kearns2010, Kearns2004, Vlamakis2013}. An emerging view of this organism holds that biofilm formation and motility (be it swimming or swarming) are carefully regulated to be mutually exclusive at the single cell level \cite{Guttenplan2013}. Biofilm formation in \textit{B. subtilis} results from the activation of several operons devoted to the formation of matrix component including exopolysaccharide (made by enzymes encoded by the \textit{eps} operon) and the protein matrix components TapA and TasA (coded by a second gene operon) \cite{Branda2006}. These two operons and others are regulated by a complex regulatory network, involving quorum-sensing \cite{OmerBendori2015}, the DegU regulator, which directly affects swarming motility \cite{Murray2009}, the Spo0A stress master regulator \cite{Chai2008} and, most clearly, the SinI-SinR-SlrR regulatory pathway \cite{Chai2010, Chai2009, Lord2019}. In this pathway, SinI acts as an inhibitor of SinR which represses biofilm formation. SinR and SlrR form a complex genetic bistable switch between biofilm formation and motile lifestyle (either swarming or swimming). The mutual exclusive nature of motility and biofilm formation is also enhanced by the \textit{epsE} gene (part of the EPS operon), whose product acts as a clutch which prevents flagellar rotation upon activation of the biofilm program \cite{Blair2008}.

Motility-induced phase separation \cite{Fily2012, Bialke2013, Cates2015} has been suggested to explain self-segregation of active particles in the absence of signaling or attracting forces. MIPS has been observed experimentally in suspensions of colloidal particles, for example Janus particles \cite{Buttinoni2013} or self-propelled rollers \cite{Geyer2019}. The main assumption underlying MIPS is that at sufficiently high densities, frequent collisions or jamming of active particles lead to a reduction in the average speed \cite{Bialke2013}. This is consistent with measurements of the average cell speed in swarm assays in which the average cell speed increases for low to intermediate densities but decreases at high ones \cite{Sokolov2007, Ariel2018}. For example, it has been speculated that MIPS underlie the fruiting body formation in \textit{Myxococcus xanthus} \cite{Liu2019} and swarming to biofilm transition in \textit{Bacillus subtilis} \cite{Grobas2021}. Recently, the standard MIPS theory have been generalized to account for elongated particle shapes \cite{Shi2018, VanDamme2019, Grossmann2020}, hydrodynamic interactions \cite{Worlitzer2021, Worlitzer2021b} and population growth \cite{Cates2010, Grafke2017}. Such additional effects can lead to arrested phase separation instead of the macroscopic phase separation expected from MIPS, i.e., finite size clusters of jammed particles, baring some resemblance to high density clusters within swarming colonies. Indeed, these theoretical studies provide further theoretical support to the MIPS triggering biofilm hypothesis \cite{Grafke2017}. 

In this paper, we present a detailed experimental study of the swarming to biofilm transition in colonies of \textit{B. subtilis}. We analyze the dynamics in the intermediate regions between the colonial center and its edge, where swarming cells coexist with dense stationary aggregates \cite{Jeckel2019}. Accordingly, we start with a quantitative analysis of the movement statistics at different stages of the swarm to biofilm transition. We show that some of the main assumptions underlying MIPS do not hold. In particular, the speed of active cells is always increasing with the local density. Next, we demonstrate that trajectories of individual cells are composed of super- and subdiffusive regimes as cells continuously enter or leave aggregates, indicating a (partially reversible) trapping scenario. In addition, experiments with biofilm-deficient mutants that do not produce any biofilm components, show that stable stationary aggregates are not possible without these components. Lastly, we use a mutant which fluoresce upon transitioning into the biofilm state to illustrate how a subpopulation of cells initiates the formation of stationary aggregates. Therefore, we conclude that a biophysical mechanism leads to trapping of motile cells, to the formation of stationary aggregates, and eventually to stable biofilms.

\section{Materials and Methods}\label{sec:exp_methods}
The bacterial background strain used in this study is \textit{Bacillus subtilis} NCIB 3610 wild-type (WT). We have used 4 variants of this strain: (i) A fluorescently labelled strain, tagged with a constitutive green fluorescent protein (GFP) (strain number 4846; sfGFP, amyE::Pveg\_R0\_sfGFP\_spec). (ii) A fluorescently labelled strain, tagged with a constitutive red fluorescent protein (RFP) (strain number 4847; pAE1222-LacA-Pveg-R0\_mKate\#2 mls). We refer to these two strain as “WT” as they behave similar to the non-tagged WT even while the fluorescence-activating light is on, with no photobleaching or changes in motility/statistics etc \cite{Peled2021}. We used these tagged strains because they are technically easier to be tracked. (iii) A WT strain coding for a transcriptional YFP reporter of the EPS promoter, which expresses YFP only upon transition into the biofilm state (strain number 1034; Peps7-3xYFP spec) \cite{OmerBendori2015}. (iv) An RFP tagged sinI mutant, defective in production of EPS and other biofilm components (strain 7871; amyE::Pveg(+1/+8)\_R0mKate2\_spec DsinI:kan). Antibiotics is added to each frozen stock to maintain the strain, also to hard-agar (2\% and 25 g/L Luria Bertani (LB)) isolation plates. Experiments were performed without additions of antibiotics. Labelling was stable.

Isolated colonies from hard-agar plates were cultured in a 2 ml LB liquid medium for each strain, and incubated at 200 rpm and 30$^\circ$C overnight. The overnight cultures were inoculated to the swarm plates. Swarm plates are standard 8.8 cm Petri-dishes filled with 15 ml of molten 0.5\% agar supplemented with LB. Swarm plates were prepared 20 hours prior to inoculation and aged at 22$^\circ$C and 40\% relative humidity, then 8 min in the flow hood. Four-$\mu$l of culture was inoculated at the center of the plates and allowed to dry for 1 hour before incubation. Swarm plates were then incubated at 30$^\circ$C until the swarm colony diameter was 4 cm and observed under the optical microscope. In order to follow trajectories of single cells in a swarm, the two WT strains, green (4846) and red (4847), were mixed, where one constitutes nearly 99\% of the population and the other 1\%. Mixing was done right prior to inoculation, in an external vial, after each one was grown separately overnight. This enabled us to track the trajectories of single cells moving among the same population, only that the color distinguishes between them.

An optical microscope (Zeiss Axio Imager Z2) equipped with LD phase contrast/fluorescence lenses of 40$\times$ and 20$\times$ was used to record microscopic swarm dynamics. The movies were captured at 50 fps and 900$\times$900 pixels for several seconds. These movies streamed directly to the hard disk, resulting in hundreds of images in a sequenced movie. Recorded movies were analysed in Python by using the Farnebäck method \cite{Farneb2003} to obtain the optical flow (OF) between consecutive frames. For details on the statistical methods refer to Appendix \ref{sec:methods}. In order to distinguish between two populations (two colors), high resolution images were taken using an Optosplit II, Andor, hooked to a Zeiss Axio Imager Z2 microscope. The system splits a dually excited image (Ex 59026x, beam splitter 69008bs, and Em 535/30; 632/60) on a NEO camera (900$\times$1800  and 50 fps) in order to generate two simultaneous but separate fields of view, green and red, that are then merged again following postprocessing (for more details see \cite{Peled2021} and Appendix \ref{sec:mth_tracking}).

\section{Results}
Following inoculation and an initial lag phase, the colony starts expanding, eventually covering the entire agar plate. The front of the colony is a highly active monolayer of swarming cells, moving rapidly in so-called rafts or dynamic clusters \cite{Jeckel2019}. Cells at the center are non-motile and are arranged in a multilayered structure, constituting a biofilm. As the colony expands, the biofilm constituting the center of the colony grows radially as well, typically much slower than the expanding front. See \cite{Jeckel2019} for a detailed  and comprehensive study of the spatio-temporal dynamics of an expanding colony of swarming {\it B. subtilis}. 

\subsection{Dynamics in the coexistence zone}

\begin{figure}
	\centering
	\includegraphics[width=\linewidth, height=\textheight,keepaspectratio]{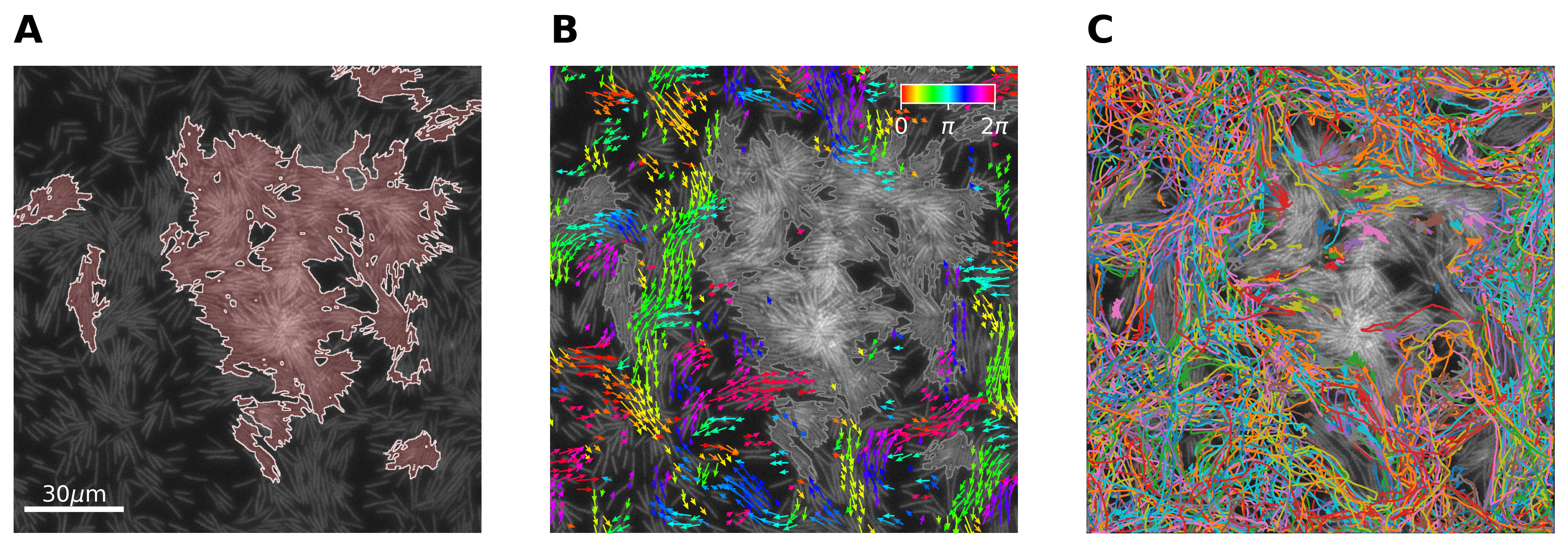}
	\caption{\textbf{Coexisting swarming and stationary cell aggregates.} \textbf{A} Stationary aggregates of cells are highlighted in red. \textbf{B} Bacterial flow field obtained from optical flow. No flow is plotted inside stationary aggregates. Color indicates orientation. \textbf{C} Individual trajectories obtained by tracking cells superimposed on microscopic image.}
	\label{fig:overview}
\end{figure}

We start by studying the swarming dynamics during the expansion phase, at intermediate regions between the colony center and its edge, where coexistence of swarming cells and stationary aggregates is found. Figure \ref{fig:overview}\textbf{A} shows stationary aggregates of cells, which are large connected regions with very low speeds. These aggregates are surrounded by swarming cells moving at significantly higher speeds. Figure \ref{fig:overview}\textbf{B} shows the coarse grained velocity of the swarming regions, obtained by optical flow analysis (see \cite{Farneb2003} and Appendix \ref{sec:mth_flow}). The distinction between the regions is a threshold speed. In addition, we analyze the trajectories of individual cells (see methods), as depicted in Fig.\ \ref{fig:overview}\textbf{C}. The trajectories reveal that cells enter and leave the stationary aggregates, intermittently. Therefore, the composition of the aggregate is not fixed and aggregate boundaries are dynamic - slowly moving, merging and splitting.

\begin{figure}
	\centering
	\includegraphics[scale=0.65,keepaspectratio]{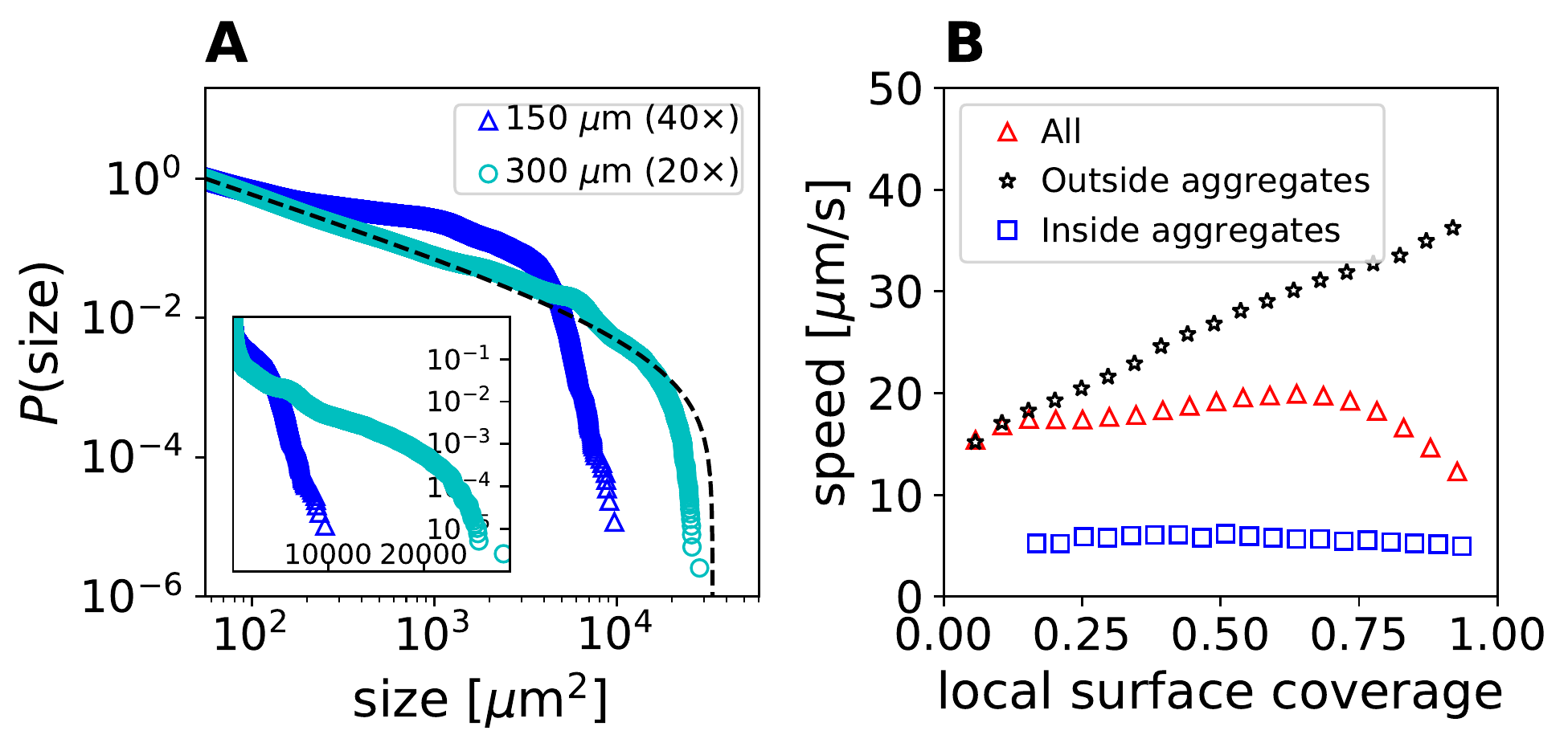}
	\caption{\textbf{Movement statistics.} \textbf{A} Aggregate size distribution. The complementary cumulative distribution $P(x)$ for fields of view with side length 150 $\mu$m (blue triangles, 40$\times$ magnification) and 300 $\mu$m (cyan circles, 20$\times$ magnification). As a visual guide, a fit to a power law with exponential cut-off at 20$\times$ is shown (dashed line). Inset shows the same data on a semilog scale. Data is obtained by averaging over several independent experiments (details in the methods and appendices). \textbf{B} Average speed as a function of {\it local} surface coverage. The speed is obtained by binning and averaging over bins with similar surface coverage. Shown is the average speed of all cells (red triangles), outside stationary aggregates (black stars) and inside stationary aggregates (blue squares). Data is obtained by averaging over several independent experiments. The global surface coverage varies between 0.5 and 0.8 for all experiments. The averaged local surface coverage inside the aggregates is on average a factor of two larger than inside.}
	\label{fig:stats}
\end{figure}

Quantitatively, Fig.\ \ref{fig:stats}\textbf{A} shows the distribution (via the empirical complementary cumulative distribution function) $P(x)$ of the aggregate sizes, i.e.\ the aggregate size distribution (ASD) at two different magnifications. The ASD appears to obey a power law over several orders of magnitude. Indeed, using a maximum-likelihood fitting and the Akaike Information Criterion (AIC), we find a power law with exponential cut-off, $P(x)=C \int_{x}^{\infty} r^{-\alpha}e^{-\lambda r}\; dr$ with $\alpha\approx1.9, \lambda\approx2\times10^{-5}$, where $C$ is a normalization constant. It is worth noting, that a power law with exponential cut-offs with similar exponets were also found to describe area size distribution of moving clusters in the “swarming” phase of several bacterial species specifically {\it B. subtilis} \cite{zhang2010} and {\it Myxococcus Xanthus} \cite{peruani2012,starruss2012} at lower densities. A simple kinetic mean-field approach reproduces such power-laws with exponential cut-off. The exponents of ASD were found to be non-universal and depend on the  loss rates of bacteria from clusters and the growth rates of clusters by fusion \cite{peruani2013}. The AIC alternatives in our analysis here were taken to be pure power laws and pure exponential distributions (Appendix \ref{sec:mth_asd}). The cut-off is related to finite size effects and depends on the size of the viewing field: At 40$\times$ magnification, the cut-off point is much lower compared to the cut-off point at 20$\times$ magnification. This indicates that larger stationary aggregates cannot be resolved by the finite observation window. Testing this hypothesis by further decreasing the magnification is technically difficult as the dynamical data needed to define stationary aggregates becomes less reliable at smaller magnifications. Overall, these findings indicate a scale free aggregate size distribution. In particular, there is no characteristic aggregate size.

Outside aggregates, cells move freely, showing coexistence of moving and non-moving cells in the same region. While previous experiments reported a non-monotonous dependence of the average speed on the density \cite{Sokolov2007, Ariel2018}, other works have argued that a monotonic reduction of speed with increased density is as an explanation for clustering of active particles \cite{Bialke2013, Cates2015}. Here, we measure the dependence of the average speed on the local surface coverage and compare the average speed in the entire viewing field with the average obtained only in- or outside the stationary aggregates. Considering the entire viewing area, Fig.\ \ref{fig:stats}\textbf{B} shows that the average speed is roughly constant for low to intermediate surface coverages, followed by a decrease at high surface coverages. This result is different than previous measurements, e.g. \cite{Beer2020,Sokolov2007, Ariel2018}, in which the average speed is significantly increasing at low to medium coverages, but decreasing at high ones. This is due to two reasons: (i) Here, samples are taken closer to the colony center, a region which has already started transforming to a biofilm. (ii) The definition of surface coverage is a bit different. Here, it refers to local coverage  (see Appendix \ref{sec:mth_binning} for details). In addition, the coverage can get very high (practically up to 1) because we are sampling local highly dense aggregates that may be starting to become multilayered. In contrast, \cite{Beer2020,Sokolov2007, Ariel2018} considered the average coverage in the entire viewing area.

Based on the distinction between swarming regions and stationary aggregates, we estimate the dependence of the average speed on surface coverage in each region separately. We find that outside stationary aggregates the average speed of freely swarming cells is always increasing with surface coverage, close to linearly. Surprisingly, this holds up to very high densities (practically 1). Again, note that speed and density are calculated locally (Appendix \ref{sec:mth_binning}), so the measurement excludes jammed aggregates.  In contrast, the speed inside stationary aggregates remains fairly constant at a low value, independent of the surface coverage. Hence, the apparent non-monotonous dependence of average speed on surface coverage for all cells is due to the different fraction of cells in the two regions. In other words, the reduction of speed at high surface coverage results from a higher fraction of cells belonging to stationary aggregates and not from slowing-down of moving cells.

\begin{figure}
	\centering
	\includegraphics[scale=.65, keepaspectratio]{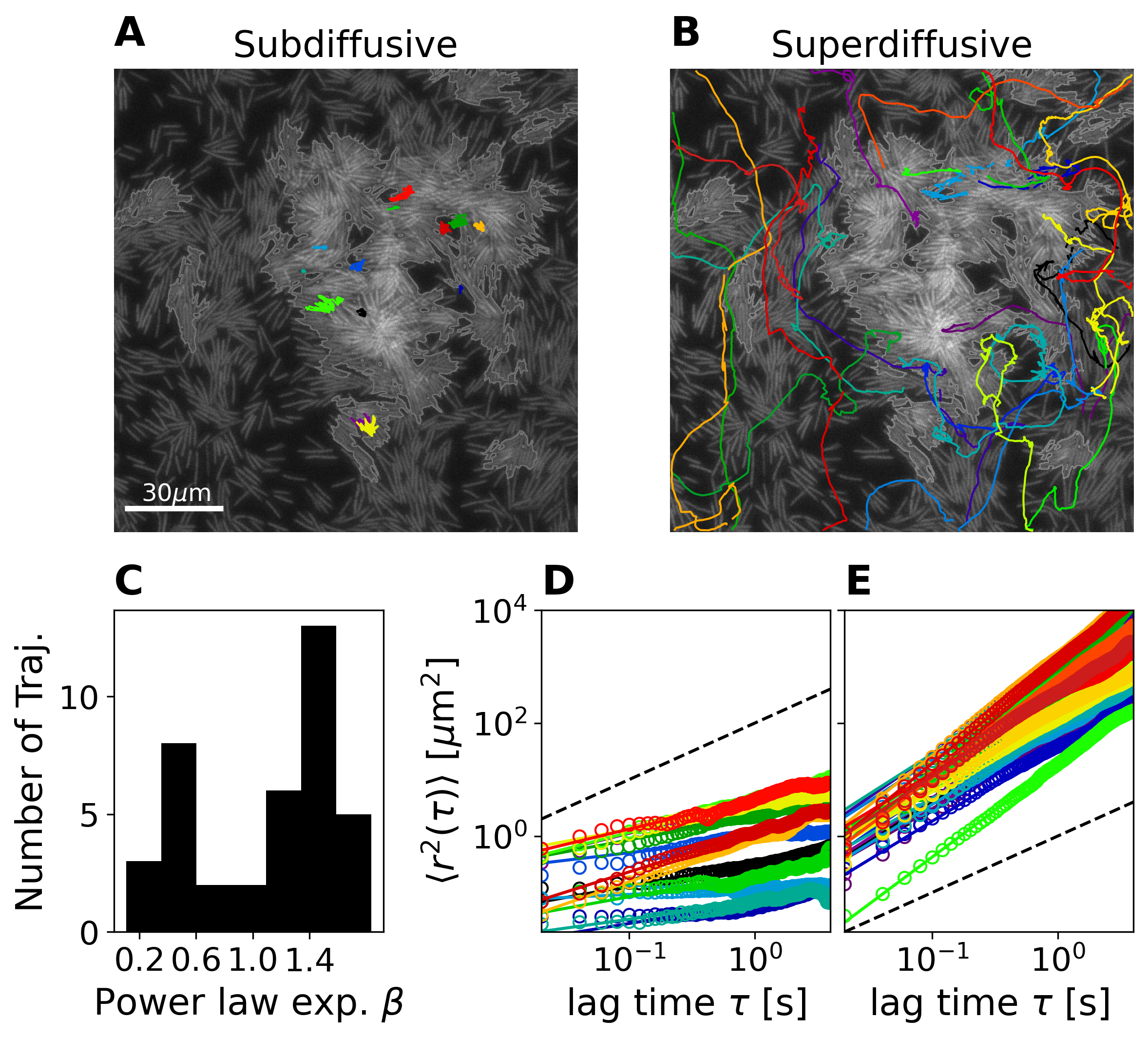}
	\caption{\textbf{Trajectory analysis.} \textbf{A} Subdiffusive and \textbf{B} superdiffusive trajectories superimposed on the microscopic image at $t=0$, while stationary aggregates are outlined. \textbf{C} Histogram of power law exponents $\beta$ obtained from non-transitive trajectories. Mean-square displacements (open circles) as a function of lag time for \textbf{D} subdiffusive trajectories (color matches the trajectories of \textbf{A}) and \textbf{E} superdiffusive trajectories (color matches the trajectories of \textbf{B}). Fits are provided as solid lines, while normal diffusion is indicated as a dashed line.}
	\label{fig:traj}
\end{figure}

On the individual scale, we study the trajectories of single bacteria within the collective, see Figs.\ \ref{fig:traj}\textbf{A} and \textbf{B}. To do so, individual cells are tracked to obtain their trajectories, denoted $r(t)$. The mean-square displacement (MSD) is computed, $\langle r^2(\tau)\rangle=\langle |r(t) - r(t+\tau)|^2\rangle$ and fitted to a power law, $\langle r^2(\tau)\rangle\sim \tau^{\beta}$. See Appendix \ref{sec:mth_tracking} for details. Figure \ref{fig:traj}\textbf{C} shows a bimodal distribution of exponents $\beta$, related to either subdiffusive ($\beta < 1$, see Fig.\ \ref{fig:traj}\textbf{D}), or superdiffusive ($\beta > 1$, see Fig.\ \ref{fig:traj}\textbf{E}) trajectories. Moreover, overlaying the trajectories with the collective motion shows that subdiffusive behaviour is found inside stationary aggregates (Fig.\ \ref{fig:traj}\textbf{A}), while superdiffusion is most commonly observed outside of stationary aggregates (Fig.\ \ref{fig:traj}\textbf{B}). A few trajectories show diffusive or transitive dynamics (i.e., sub- or superdiffusive on different time scales). Manual inspection showed that such trajectories correspond to bacteria that entered or left aggregates during the periods they were tracked. 

Overall, the analysis of trajectories suggests the following microscopic picture: Within the swarming regions, cells are superdiffusive, in accordance with previous results, showing that trajectories of individual swarming cells are consistent with L\'evy walks \cite{Ariel2015,Ariel2017}. In contrast, cells trapped inside stationary aggregates show subdiffusive behaviour. Furthermore, there is a slow exchange of cells between the swarm and the aggregates, as observed in transitive trajectories. That is, swarming cells can get trapped in stationary aggregates, while trapped cells can also leave a stationary aggregate and resume swarming.

Next, we study the mechanism leading to the nucleation of scale-free stationary aggregates. The linear increase of speed with surface coverage for swarming cells indicates that stationary aggregates do not emerge due to jamming at high densities. However, trajectory analysis suggest that trapping is partially responsible to the creation and dynamics of aggregates. We will now show that these observations are related to the biological cellular processes of cells initiating the biofilm program.

\subsection{Biophysical origin of the stationary aggregates}
\begin{figure}
	\centering
	\includegraphics[width=\linewidth, height=\textheight, keepaspectratio]{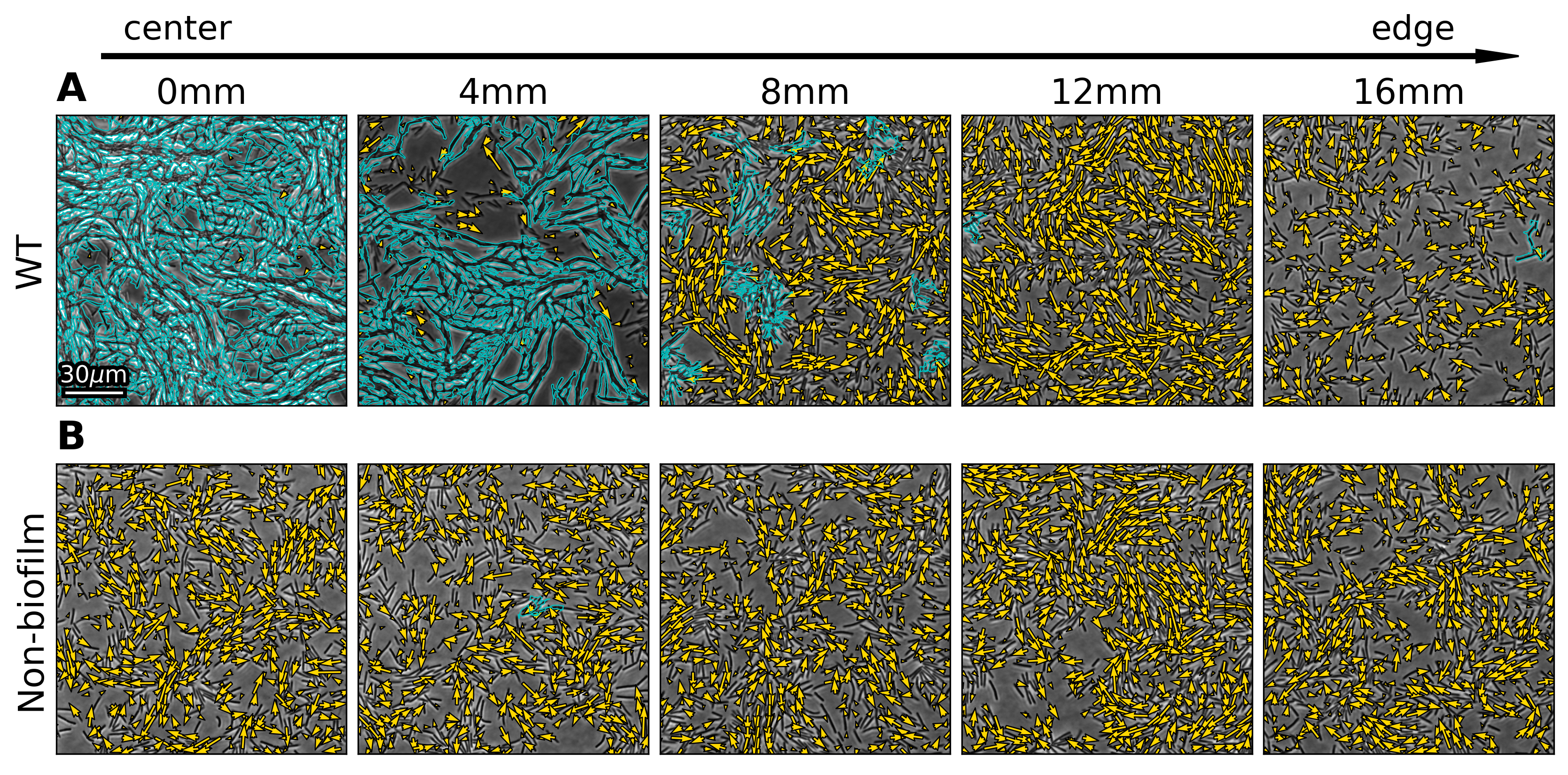}
	\caption{\textbf{Wild-type vs.\ non-biofilm mutant.} Snapshots of \textbf{A} the WT and \textbf{B} the non-biofilm mutant for a variety of regions from the edge of the inoculum (center) towards the edge of the colony in steps of 4mm (left to right). Flow is superimposed as in Fig.\ \ref{fig:overview}\textbf{B}, while stationary aggregates, detected using optical flow analysis, are encircled in cyan.}
	\label{fig:mutant_snapshot}
\end{figure}

To test the hypothesis that the presence of stationary aggregates is related to biofilm formation, we use a mutant that is defective in the production of biofilms (strain 7871; see Methods). Hence, we will refer to the mutant as 'non-biofilm mutant'. In particular, the mutant cannot produce extracellular polymeric substances (EPS), among other things. Figure \ref{fig:mutant_snapshot} shows typical snapshots of the dynamics for a wild-type (WT) colony and for the non-biofilm mutant at different distances from the colony center (edge of the inoculum). In the WT colony, the center is dominated by densely packed, stationary cells that form a biofilm. Towards the edge of the WT colony, swarming is observed, while in between the center and the edge, swarming cells and stationary aggregates coexist. The non-biofilm mutant shows uniform swarming behaviour throughout the entire colony. That is, the dynamics looks similar independent of the location within the colony. Note that the optical flow algorithm detects a few spurious small stationary aggregates, that are probably caused by cell jamming. However, these aggregates do not grow to the large, scale-free aggregates observed for the WT.

\begin{figure}
	\centering
	\includegraphics[scale=0.7, keepaspectratio]{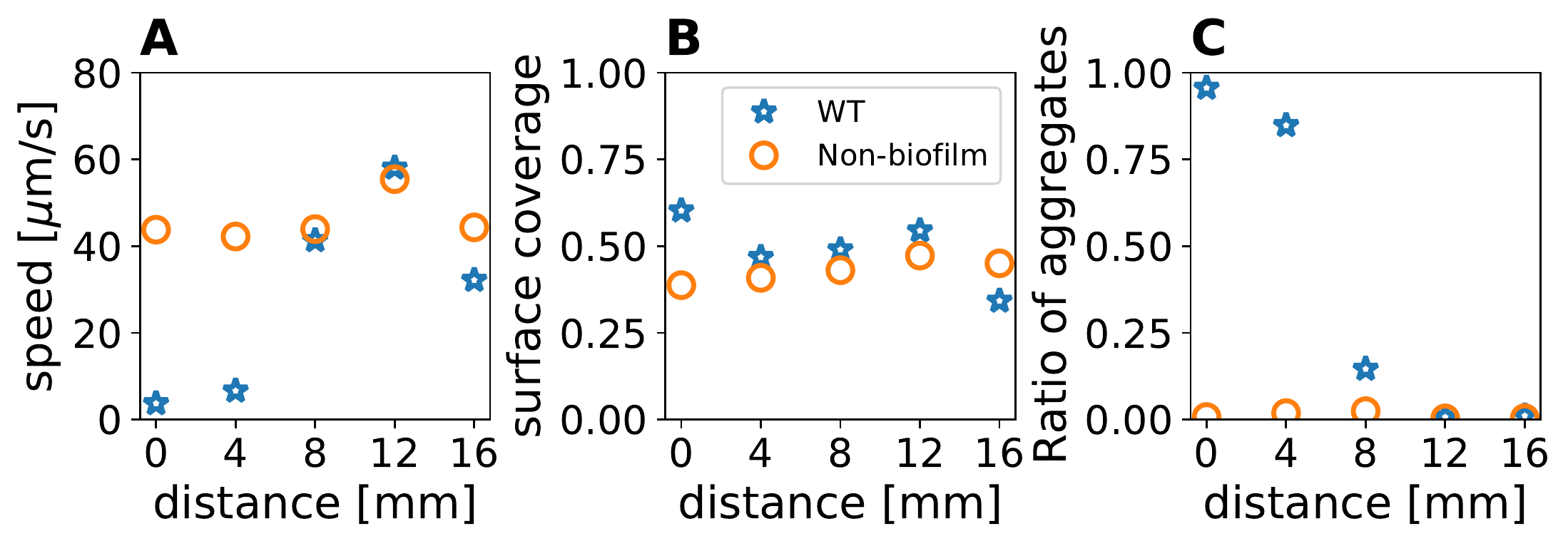}
	\caption{\textbf{Wild-type vs.\ non-biofilm mutant: Statistics.} \textbf{A} Average speed, \textbf{B} surface coverage and \textbf{C} ratio of aggregates for the WT (blue stars) and the non-biofilm mutant (orange circles) at different distances from the edge of the inoculum.}
	\label{fig:mutant_statistics}
\end{figure}

To quantify the differences between the WT and the non-biofilm mutant, we measure the average speed, surface coverage and the ratio of stationary aggregates at different distances from the center.
Following \cite{Jeckel2019}, studying different spatial regions is equivalent to looking at different times during the colonies expansion, where looking close to the inoculation center is analogous to later times. The ratio of aggregates is defined as the ratio of cells which belong to a stationary aggregate to all cells in the field of view. For the non-biofilm mutant, there are almost no stationary aggregates, and the speed and density remain fairly constant, see Fig.\ \ref{fig:mutant_statistics}. For the WT, there are clear differences between the center and the edge of the colony. Close to the center, surface coverage is high and almost all cells are inside stationary aggregates. Accordingly, the average speed is very low. Towards the colony edge, the statistics are comparable to the non-biofilm mutant. We remark that the difference in speed for the last data point between the WT and the non-biofilm mutant is correlated with the difference in surface coverage. Hence, the slightly slower WT speeds can be explained by the slightly lower densities, compare Fig.\ \ref{fig:stats}\textbf{B} and \cite{Beer2020}. Concluding, the main difference between the WT and the non-biofilm mutant is the absence/presence of large stationary aggregates. This supports the hypothesis that the stationary aggregates are related to biofilm formation.

\begin{figure}
	\centering
	\includegraphics[scale=0.5, keepaspectratio]{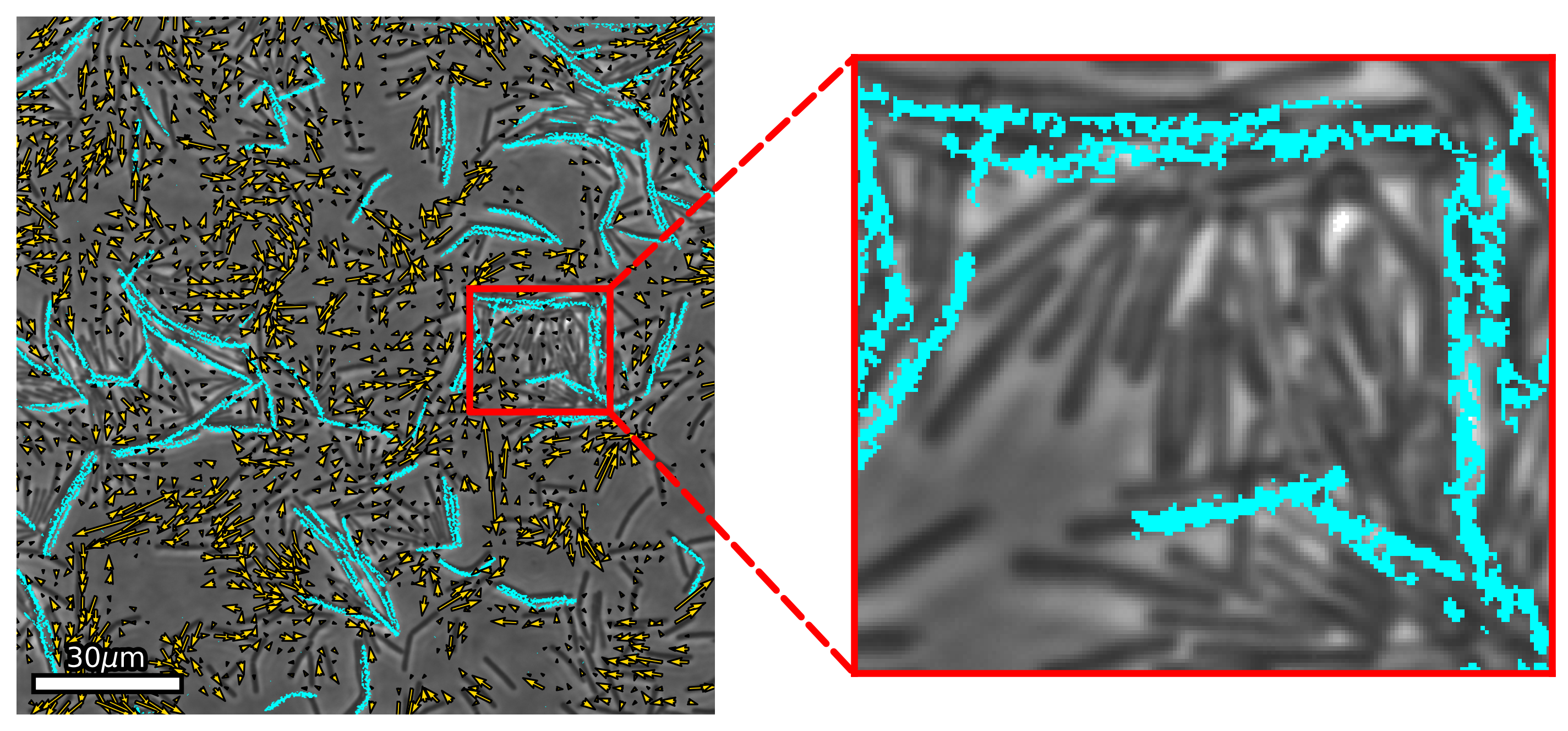}
	\caption{\textbf{Nucleation of stationary aggregates.} Snapshots of stationary aggregates. Cells secreting EPS are highlighted in cyan. Bacterial flow field is provided by yellow arrows. Zoom-in highlights the trapping of motile cells by EPS-secreting cells.}
	\label{fig:EPS_tracers}
\end{figure}

It is well-known \cite{Kearns2005,Jeckel2019} that genetically identical cells within a colony of \textit{B. subtilis} show phenotypic differentiation. In particular, only a fraction of cells, the so-called matrix-builders \cite{Chai2008}, produces EPS, increasing adherence to surfaces and between cells. This suggests EPS as a possible trapping mechanism that nucleates the stationary aggregates. To illustrate this point, we use strain expressing YFP upon secretion of EPS (strain 1034; see Methods), see Fig.\ \ref{fig:EPS_tracers}. Indeed, we observe that cells secreting EPS are immotile. As predicted by simple models of self-propelled particles \cite{Kaiser2012}, such stationary obstacles can act as particle traps and become seeds for stationary aggregates. Motile cells may push around immotile cells, for example if sufficiently many cells align to form a moving cluster. Again, such a behaviour has been predicted from theory \cite{Pietzonka2019}. Altogether, we conclude that mechanical and biological mechanisms combine to form stationary aggregates: A sub-population of cells in the colony are immotile matrix-producers. Depending on several factors, for example the arrangement and alignment of cells or the strength of cell-substrate adhesion among others, immotile cells either get rearranged by motile cells or form traps, which nucleate stationary aggregates.

\section{Discussion}
Analyzing the dynamics in different zones between the colony edge and its center by means of statistical physics, allows us to identify some of the fundamental biophysical mechanisms underlying the formation of stationary bacterial aggregates and eventually biofilms. We find that a few cells initiating the biofilm program act as seeds for stationary aggregates. Physical interactions (e.g., mechanical or steric) lead to trapping of swarming cells, and to growth of aggregates with scale-free size distributions. At later stages, most cells within the aggregates transition to biofilm as well.
Importantly, we found that {\it macroscopic phase separation} (or arrested aggregation with scale-free size distributions) is not observed without cells transitioning into biofilm. Hence, motility induced phase separation does not play a role within swarming colonies. This does not mean that density fluctuations, leading to {\it transient and local} creation of small aggregates of temporarily immobile cells are not an important step in triggering the biofilm program in the initial stages of the swarm to biofilm transition.

The biological advantages of the stationary aggregates remain unclear. We suggest two possible scenarios: 1) Stationary aggregates are early form/precursors of biofilms. During the spread of the colony through swarming, a sub-population of cells responds to environmental cues, producing EPS. These cells act as founder cells for new biofilms far away from the inoculum. 2) Stationary aggregates accelerate the spreading of biofilms. A colony of \textit{B. subilis} intrinsically features sub-populations of swarming cells and EPS secreting cells. Due to steric interactions, EPS producing cells are transported away from the inoculum, where they nucleate stationary aggregates. In either scenario, stationary aggregates help to colonize the surface with biofilms. Expansion of biofilms is usually mediated by the growth and division of individual cells within the biofilm. This process is slow, leading to limited exploration of the available space. On the other hand, swarming facilitates rapid expansion and surface exploration. Hence, the formation of stationary aggregates may accelerate the spread of biofilms over surfaces on much shorter time scales compared to growth mediated expansion. The latter scenario questions the nature of swarming, as it might be a temporary state whose main role is to accelerate the spread of biofilms. Hence, swarming and biofilms are not antagonistic states, but evolutionary beneficial to ensure a rapid spreading of sheltered communities in the form of biofilms.

{\bf Acknowledgements}
AE and IG were funded by European Research Council grant no. 724805. VMW, AB, GA, MB and SH thank partial support from the Deutsche Forschungsgemeinschaft (The German Research Foundation DFG) Grant No. HE5995/3–1 and Grant No. BA1222/7–1.

\normalem
\bibliography{references}
\bibliographystyle{unsrt}

\clearpage
\appendix

\section{Data Acquisition}
The data for Fig.\ \ref{fig:stats} is obtained from five individually recorded movies in the transition zone. All movies were captured within $\approx$ 5 minutes. Three movies were taken at 40$\times$ magnification and two movies at 20x magnification. One movie of each magnification is recorded for 5000 frames, whereas the rest is recorded for 2000 frames each. Movies were captured at 50 frames per second.

\section{Data Analysis}\label{sec:methods}

\subsection{Flow computation and aggregate identification}\label{sec:mth_flow}
\begin{figure}
	\centering
	\includegraphics[width=\linewidth, height=\textheight,keepaspectratio]{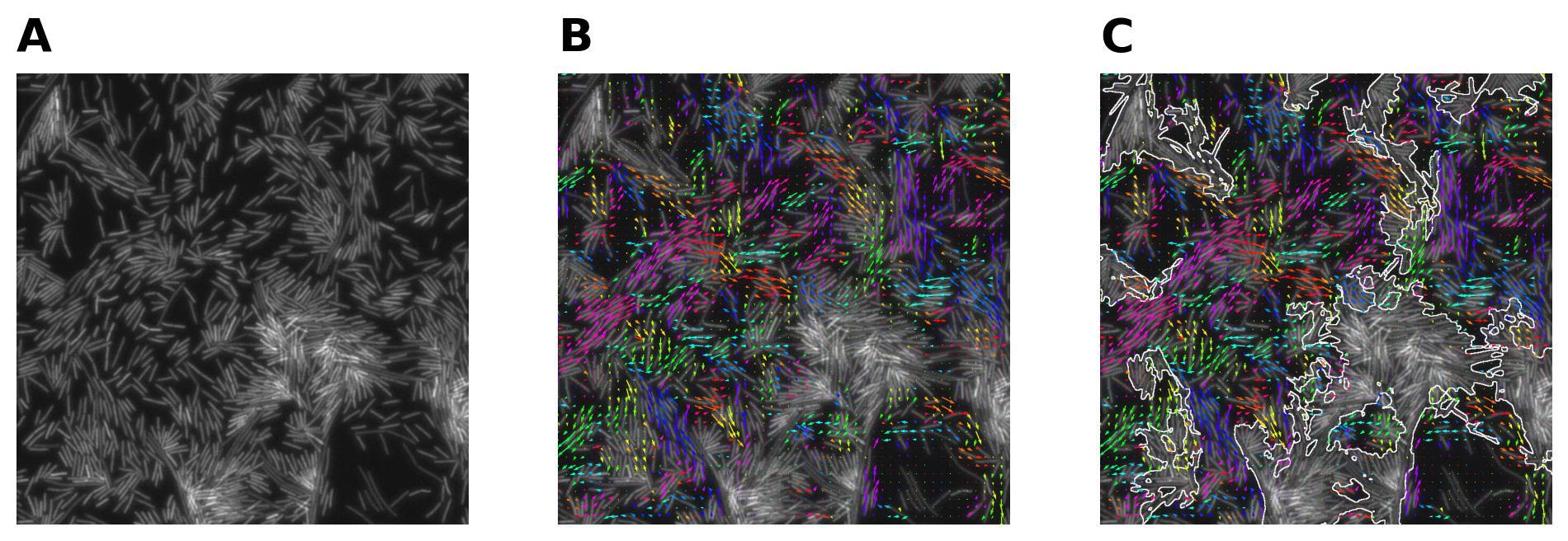}
	\caption{\textbf{Identification of stationary aggregates.} \textbf{A} Raw image obtained from the microscope. \textbf {B} Computed flow superimposed on the raw image. Color code of the arrows represent the angle of the flow, i.e.\ areas with the same color show cells moving in the same direction. \textbf{C} Stationary aggregates, i.e.\ slow areas of at least 10 bacterial surface coverages, are encircled and plotted along the flow.}
	\label{fig:algo}
\end{figure}

The raw image data obtained from the experiments is used to compute the flow and identify different regions as illustrated in Fig.\ \ref{fig:algo}. First, the optical flow is computed via the Farnebäck algorithm \cite{Farneb2003}, which provides the local velocity field $\mathbf{v}(\mathbf{x},t)$ shown in Fig.\ \ref{fig:algo}\textbf{B}. Parameters of the Farnebäck algorithm are provided in table \ref{table:of}. We then identify stationary clusters in four steps: First, the original image is smoothed once with a Gaussian kernel. Then, cells are separated from the background by thresholding the smoothed image. Based on the flow, nearly stationary regions are identified (i.e. regions below a certain magnitude of the flow). Finally, connected stationary regions above a certain size are labeled as a stationary cluster, see Fig.\ \ref{fig:algo}\textbf{C}. We require stationary aggregates to at least cover the same area as roughly ten bacteria. From our data, we estimate the surface coverage of a typical cell to be roughly 200 pixels (at 40$\times$ magnification), hence we set a size threshold of 2000 pixels (at 40$\times$ magnification).

\begin{table}[h!]
\centering
\begin{tabular}{|| c c c||} 
 \hline
 Parameter & 40$\times$ & 20$\times$ \\
 \hline\hline
 pyr\_scale & 0.5 & 0.5 \\ 
 levels & 3 & 3\\
 winsize & 20 & 10 \\
 iterations & 3 & 3\\
 poly\_n & 5 & 5\\
 poly\_sigma & 1.1 & 1.1\\
 flags & 0 & 0\\
 \hline
\end{tabular}
\caption{Parameters used for the Farnebäck algorithm \cite{Farneb2003} as implemented in OpenCV \cite{opencv_library} for 20$\times$ and 40$\times$ magnification.}
\label{table:of}
\end{table}

\begin{table}[h!]
\centering
\begin{tabular}{|| c c c||} 
 \hline
 Parameter & 40$\times$ & 20$\times$ \\
 \hline\hline
 $\sigma$ & 2 & 1 \\ 
 min\_area & 2000 & 500\\
 min\_speed & 2 & 1 \\
 \hline
\end{tabular}
\caption{Parameters used to identify stationary aggregates, i.e.\ $\sigma$ (size of the Gaussian kernel used for smoothing), min\_area (minimal size of an aggregate in pixels), min\_speed (speed threshold for aggregates in pixels per frame).}
\label{table:aggregate_parameter}
\end{table}

\subsection{Binning}\label{sec:mth_binning}
\begin{figure}
	\centering
	\includegraphics[width=\linewidth, height=\textheight,keepaspectratio]{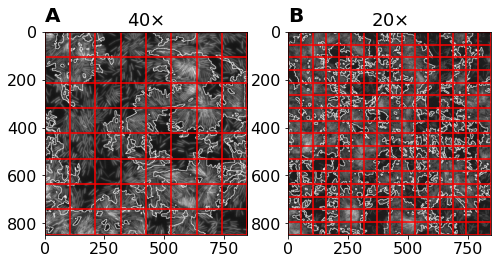}
	\caption{\textbf{Binning approach.} Bins superimposed on microscopic image with otlined stationary aggregates at \textbf{A} 40$\times$ magnification and \textbf{B} 20$\times$ magnification.}
	\label{fig:binning}
\end{figure}

To estimate quantities such as the local velocity and speed, we divide each snapshot into 8$\times$8 (40$\times$ magnification) or 16$\times$16 (20$\times$ magnification) bins, see Fig.\ \ref{fig:binning}. For each bin, the surface coverage as well as the average speed of cells is measured. Due to the large size of the bins, we averaged the resulting data in Fig.\ \ref{fig:stats}\textbf{A} over bins with similar surface coverage to reduce fluctuations. To obtain the statistics inside and outside stationary aggregates (see Fig.\ \ref{fig:stats}\textbf{A}), we either neglect all bins that contain stationary aggregates or only consider bins in which all cells are part of a stationary aggregate respectively.

\subsection{Aggregate size distribution}\label{sec:mth_asd}
We assume that the aggregate size distribution (ASD) shown in Fig.\ref{fig:stats}\textbf{B} can be reasonably well approximated by a statistical model. To compare the relative quality of different models, we use the Akaike information criterion (AIC). Our choice of models is motivated by the empirical complementary cumulative distribution function of the data. In particular, we consider an exponential distribution
\begin{equation}
    f_\lambda(x)=\lambda e^{\lambda(x_{min}-x)},
\end{equation}
a power law
\begin{equation}
    f_\alpha(x) = \frac{\alpha-1}{x_{min}}\left(\frac{x}{x_{min}}\right)^{-\alpha},
\end{equation}
and a power law with exponential cut-off
\begin{equation}
    f_{\alpha,\lambda}(x) = C(x_{min},\lambda,\alpha)x^{-\alpha}e^{-\lambda x}
\end{equation}
with
\begin{equation}
    C(x_{min},\lambda,\alpha) = \frac{\lambda^{1-\alpha}}{\Gamma(1-\alpha,\lambda x_{min})},
\end{equation}
where $\Gamma(s,x)=\int_{x}^{\infty}r^{s-1}e^{-r}dr$ is the upper incomplete gamma function. Note that one can obtain an exponential distribution and a power law from the the power law with exponential cut-off in the limits $\alpha\to0$ and $\lambda\to0$ respectively. The minimal value $x_{min}$ is given by the threshold value we used to identify aggregates, see Appendix \ref{sec:mth_flow}. That is, we assume that the statistical model holds for the entire range of the data. Estimates for the parameters $\lambda$ and $\alpha$ are obtained from a maximum likelihood approach.

Computing the AIC for the data recorded at 20$\times$ magnification for the three candidate models reveals that the power law with exponential cut-off is several times more likely to produce the data than the alternatives. In Fig.\ref{fig:stats}\textbf{B} the tail distribution of a power law with exponential cut-off, i.e.\ $P(x) = \int_{x}^{\infty}f_{\alpha,\lambda}(r)\; dr$, is shown as a visual guide.

\subsection{Single-cell tracking}\label{sec:mth_tracking}
\begin{figure}
    \includegraphics[width=\linewidth, height=\textheight,keepaspectratio]{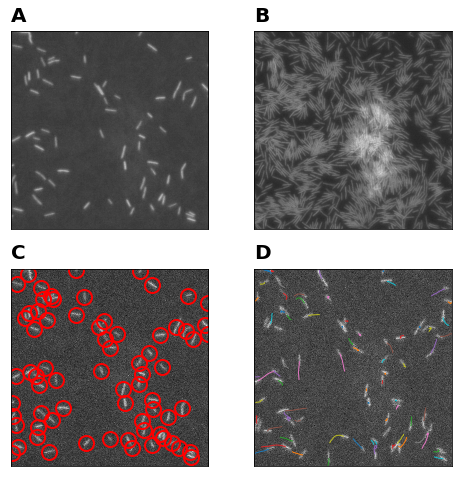}
    \caption{\textbf{Single-cell tracking.} \textbf{A} Left panel (see text), representing roughly 1\% of all cells within the field of view. \textbf{B} Right panel (see text), representing the exact field of view at the same time and the remaining 99\% of all cells within the field of view. \textbf{C} Identified cells from the left panel are encircled in red. \textbf{D} Trajectories of length $0.2$s obtained from the left panel superimposed on the microscopic image at $t=0.2$s. Each trajectory is represented by an individual color.}
	\label{fig:methods_traj}	
\end{figure}

To track single cells as well as the collective behavior simultaneously, the colony was prepared as described in section \ref{sec:exp_methods}. This procedure results in two images, obtained at the same position within the colony at the same time, where one image (left panel, see Fig.\ \ref{fig:methods_traj}\textbf{A}) shows roughly 1\% of the cells and the other image (right panel, see Fig.\ \ref{fig:methods_traj}\textbf{B}) the remaining 99\%. Hence, the position of the few cells in the left panel within the collective (right panel) can be obtained by merging the images. Trajectories of cells in the left panel were obtained by identifying cells via Hessian- and Frangi-filters (see Fig.\ \ref{fig:methods_traj}\textbf{C}) and using the Crocker–Grier algorithm \cite{Crocker1996} implemented in trackpy \cite{trackpy} (see Fig.\ \ref{fig:methods_traj}\textbf{D}).

We compute mean-square displacements (MSD) of individual trajectories $\langle r^2(\tau)\rangle=\langle |r(t) - r(t+\tau)|^2\rangle$, where $r(t)$ is the position of a cell at time $t$ and the average is taken over all suitable times $t$. For large times one expects the relation $\langle r^2(\tau)\rangle\sim \tau^{\alpha}$, where $\alpha$ indicates diffusive ($\alpha\approx 1$), subdiffusive ($\alpha\ll1$) or superdiffusive ($\alpha\gg 1$) behaviour. Based on a fit of the MSD to a power law, we classify trajectories into the aforementioned categories. Furthermore, we introduce a transitive category to account for trajectories which fit poorly to a power law, i.e.\ if the standard deviation of the fit of $\alpha$ is large. When fitting to an power law, we log transform the $\tau$ and $\langle r^2(\tau)\rangle$ data and use linear regression to get the best linear approximation.

\end{document}